# Quantification of Coulomb Interactions in layered Lithium and Sodium Battery Cathode Materials

Bongjae Kim[1], Kyoo Kim[2], Sooran Kim[3*]

[1]Department of Physics, Kunsan National University, Gunsan 54150, Korea
[2]Korea Atomic Energy Research Institute (KAERI), 111 Daedeok-daero, Daejeon 34057, Korea
[3]Department of Physics Education, Kyungpook National University, Daegu 41566, Korea

**ABSTRACT:** Despite the importance of the electron correlation in the first-principles description of the Li-ion cathode materials, the Coulomb interaction parameter, $U$ is often treated as an *ad hoc* value. In practice, one usually relies on empirical ways of parametric treatment of $U$ to optimally match the experimentally observed physical properties such as band gap or reaction energy. Here, using constrained random phase approximation (cRPA) method, we self-consistently evaluate the Coulomb $U$ and Hund $J$ values for representative layered cathode materials including not only Li compounds but also Na compounds; $LiCoO_2$, $LiNiO_2$, $LiMnO_2$, $NaCoO_2$, $NaNiO_2$, and $NaMnO_2$. We found that the Coulomb interaction parameters for Li and Na compounds and their polymorphs with different layer stackings do not deviate much, which shows the dominant role of local environment rather than of global structural features. We have analyzed the origin of variable Coulomb parameters, which is mainly due to the competition between the localization and screening. We provided cRPA Coulomb parameters for battery cathode materials and validate the values by observing systematic improvement in describing the experimentally observed average intercalation voltage and lattice parameters. These results can be applied for the first-principles calculations as well as model-based simulations for the theoretical investigation of cathode systems.

## 1. INTRODUCTION

First principles calculation or *ab initio* calculation based on the density functional theory (DFT) is a widely used theoretical method for both academic and industrial applications. Being an *ab initio* scheme, the DFT is a parameter-free approach, which can explain the ground state properties of real materials without any adjustable physical parameters. This is a great advantage over other theoretical methods, which requires specific physical parameters, usually from experiments or judicious guessing as inputs for numerical calculations. For rechargeable battery materials, the DFT has established itself as a representative theoretical approach in the description of physical properties such as ionic diffusivity, intercalation voltage, and formation energy by its ability of realistic description of the electronic structures. In fact, both the model-based approaches and the DFT were instrumental in the field of the cathode materials by complementing each other. In the development of the original idea of cathode systems, the simplistic model provided valuable concepts and insight by describing the working mechanisms with a few key parameters such as charge-transfer energy, bandwidth, and Coulomb interaction.[1–3] Along with that, the DFT offers realistic demonstrations of materials from the first-principles.[4–8]

Despite its common use, being a mean-field approach, the conventional DFT calculation often fails in describing the so-called correlated system, such as transition metal oxides (TMOs). This is originated from the strongly correlated nature of transition metal *d*-orbitals, which governs the low-energy dynamics of the system. Introducing a tunable Hubbard $U$ parameter (DFT+$U$) significantly improves the reproducibility of calculations for experimental physical properties, which comes at the expense of one of the most important merits of the DFT - being the first-principles method. Accordingly, the treatment of correlated *d*-orbitals within the first-principles scheme has been a longstanding issue in the theoretical condensed matter community.[9]

As most of the cathode materials belong to the category of TMOs, the same issue should naturally arise. For example, it has been known that DFT calculations underestimate the intercalation voltages of Li-ion cathode materials.[10–13] The common treatment is simply adding parametric Coulomb interaction $U$ within the DFT+$U$ formalism[10–15] to match the experimental intercalation voltage. However the calculated voltage can vary significantly upon different $U$ values[10,11] which weaken the DFT's predictive power. Therefore, the correct estimation of the Coulomb interaction $U$ is of paramount importance for the coherent understanding of cathode materials as well as for the reasonable prediction of material properties.

The $U$ parameter can be determined to fit the experimental values or self-consistently evaluated. Most widely accepted $U$ parameters set for battery materials these days are obtained by fitting the experimental reaction energies.[16] On the other hand, several methods which determines the Coulomb interaction parameters within DFT scheme have been developed.[17–20] For Li-ion cathode materials, there have been previous studies of Coulomb parameters based on the linear response method.[10,21,22]. The constrained random phase approximation (cRPA) is another approach that can directly calculate the Coulomb interaction matrix from the one-particle and two-particle parts of the Hamiltonian on an equal footing. The cRPA method is widely used in close connection to the advanced many-body approach such as dynamical mean field theory (DMFT) and beyond.

Compared to other schemes, the cRPA has a few useful characteristics. One can identify the detailed screening channels of the specific orbital and analyze the Coulomb interaction parameters $U$ and Hund $J$, from the microscopic point of view. The extension to advanced methods employing such as frequency-dependent $U(\omega)$ is simple, which is expected to be the next step beyond DFT+$U$ and DMFT.[9,23,24] Most importantly, the cooperation with a model-based study is straightforward, which can help to identify the basic principles and working mechanisms of the system – this will eventually establish the designing principle of new cathode materials.

In this study, we systematically investigate the Coulomb interaction parameters employing the cRPA approach for various layered cathode materials of Co, Ni, and Mn compounds. We evaluate and compare $U$ values for various cases not only intercalated and de-intercalated structures of Li/Na compounds, but also their several polymorphs with different layer stacking. The origin and effects of Coulomb $U$ variation upon different phases are analyzed using the cRPA Hubbard parameters and the density of states of transition metal element and oxygen. To validate our approach, we calculate the average intercalation voltage and lattice parameters for Li and Na compounds and compared the results with previous experimental data.

## 2. METHODS

**2.1 cRPA calculation.** The Coulomb interaction parameters are quantified fully *ab initio* by the cRPA method. As a first step, we choose the correlated target bands, in this case, the transition metal $d$-orbitals. Secondly, the contribution of screening within the target bands, $P^c$ is removed from the total polarizability $P$; $P^r = P - P^c$. Thirdly, the partially screened Coulomb interaction kernel, $U$, is calculated by solving the following equation:

$$U^{-1} = [U^{bare}]^{-1} - P^r$$

where the $U^{bare}$ is unscreened Coulomb interaction kernel. The final matrix elements of $U$ are evaluated by

$$\mathbf{U}_{ijkl} = \lim_{\omega \to \infty} \iint d^3r d^3r' \omega_i^*(\mathbf{r}) \omega_k^*(\mathbf{r}')\mathbf{U}(\mathbf{r},\mathbf{r}',\omega)\omega_j(\mathbf{r})\omega_l(\mathbf{r}')$$

where $\omega_j(\mathbf{r})$s are the maximally localized Wannier functions obtained by the Wannier90 code[25–27] which serve as local basis functions, and $\mathbf{U}(\mathbf{r},\mathbf{r}',\omega)$ is the frequency-dependent partially screened interaction kernel. The detailed procedure of our cRPA method can be found in the previous reports[28,29]. For simplicity, in our work, the resulting Coulomb $U$ and the Hund's coupling parameter $J$ for the $d$-orbitals are obtained by averaging $\mathbf{U}_{ijij}$ and $\mathbf{U}_{ijji}$ matrix elements. The so-called $d/d$-$p$ model is used to account for the effect of O-$p$, and to mimic the DFT+$U$ implementation on atomic orbitals by maximizing the local character of the Wannier function.

**2.2 DFT calculation** All DFT calculations were performed using the Vienna ab initio simulation package (VASP).[30] We utilized three exchange-correlation functionals: Perdew-Burke-Ernzerhof (PBE)[31] and newly developed SCAN (Strongly constrained and appropriately normed semilocal density functional) functional.[32,33] We further included van der Waals (vdW) correction using the zero damping DFT-D3 method of Grimme.[34] DFT+$U$ method was used to consider the correlated $d$ orbitals for transition metal ions within the simplified rotationally invariant Dudarev method.[35] The energy cut for the plane waves and $k$-point density were 650 eV and 5000/atom, respectively. The structural optimization including lattice parameters and atomic positions within spin-polarized DFT calculations were performed.

Figure 1 shows the structures we choose. $LiCoO_2$ and $NaCoO_2$ are crystallized in hexagonal $R\bar{3}m$ structure (O3 structure).[36,37] Monoclinic $LiNiO_2$, $NaNiO_2$, $LiMnO_2$, and $NaMnO_2$ are selected with a space group of $C2/m$ (O'3 structure).[38–41] De-intercalated structures of $CoO_2$, $NiO_2$, and $MnO_2$ are considered in the hexagonal structure with a space group of $P\bar{3}m1$, which corresponds to O1 structure.[42–46] We also tested representative polymorphs: O3-$LiNiO_2$[47], P2-$NaCoO_2$[48], P'2-$NaMnO_2$[49], and O'3-$NiO_2$.[42] In O3 type and P2 type structures, alkali ions are located at octahedral and prismatic sites, respectively, and the number, 3 or 2 represents the number of different transition metal layers. The prime symbol indicates structures containing the in-plane Jahn-Teller distortion.

When alkali ion insertion reaction is written by the equation, $A_{x2}B \leftrightarrow A_{x1}B + (x_2 - x_1)A$, the average intercalation voltage $\langle V \rangle$ or redox potential for alkali ion insertion can be calculated by the following equation with $x_2 > x_1$[12,50]:

$$\langle V \rangle = -\frac{[E(A_{x2}B) - E(A_{x1}B) - (x_2 - x_1)E(A)]}{(x_2 - x_1)e}$$

where $E(A_{x2}B), E(A_{x1}B)$, and $E(A)$ are the total energy/formula unit (f.u.) of structures of the alkali-ion intercalation, de-intercalation, and the bcc Li or Na, respectively. $e$ represents the value of the electron charge.

## 3. RESULTS AND DISCUSSIONS

In the DFT+$U$ calculation, the Coulomb correlation of localized orbitals, for instance in $d$-orbitals for TMOs, is explicitly treated in the form of the Hubbard-type model. Since the Coulomb $U$ parameter depends on the inverse of the distance between point charges, the electron localization, here represented by the Wannier spread, is a measure of bare $U$ values. However, the bare Coulomb potential in real material with many other electrons is suppressed by polarizability screening $P^r$, through the hybridization with extended orbitals, mainly by O-$p$ orbitals in our systems, and the detailed energy distribution of orbitals, for instance by crystal field, and so on. Therefore, resulting screened $U$ value is the outcome of complicated mechanisms by the interplay of localization, screening, and structural peculiarities.

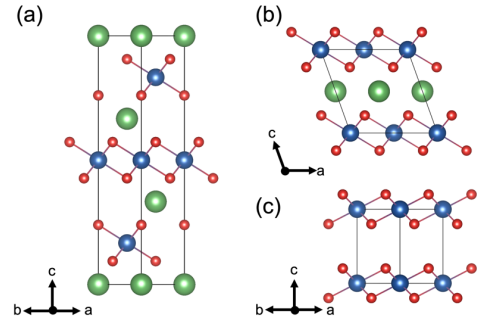

**Figure 1.** Structures of (a) $ACoO_2$ ($A$=Li, Na) in the $R\bar{3}m$ space group (O3 structure), (b) $ANiO_2$ and $AMnO_2$ in the $C2/m$ space group (O'3 structure), (c) $CoO_2$, $NiO_2$, and $MnO_2$ in the $P\bar{3}m1$ space group (O1 structure). The green, blue, and red balls indicate Li/Na, Co/Ni/Mn, and O, respectively.

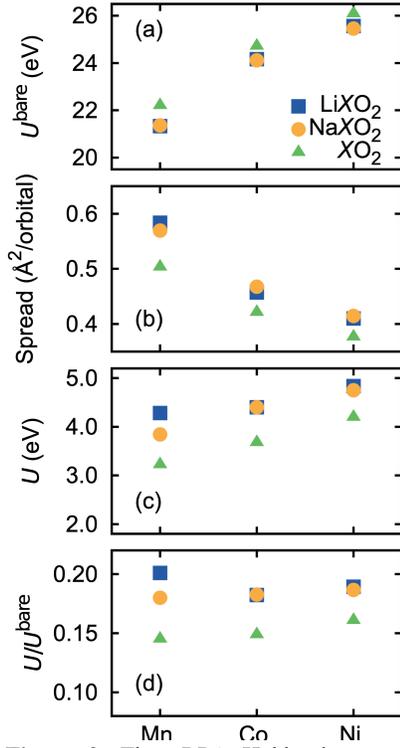

**Figure 2.** The cRPA Hubbard parameters; (a) $U^{bare}$, (b) Wannier spread, (c) $U$ and (d) $U^{bare}/U$ in Mn, Co, and Ni systems with different alkali ion intercalations.

Figure 2 shows the results of the cRPA calculations for the representative Mn, Co, and Ni-based cathode systems. Li- and Na compounds with a same TM element have similar Hubbard parameters. In both $AXO_2$ and $XO_2$ cases, as the electron occupancy increases from Mn- to Ni-based systems, the $U^{bare}$ value in Fig. 2(a) systematically increases as expected for general TMOs. This is due to the enhanced localization of $d$-orbitals upon occupancy, which is demonstrated with systematic decreases in the averaged Wannier spread per each $d$-orbital of correlated orbitals as shown in Fig. 2(b). This trend is consistent with that in other TMO systems.[51,52]

We note here that among the same TM systems, the $U^{bare}$s for $XO_2$ (X: Mn, Co, Ni) systems are slightly larger than those of Li/Na$XO_2$ systems. Specifically, the $U^{bare}$s of TM-$d$ ion for $MnO_2$, $CoO_2$, and $NiO_2$ is 22.2, 24.7, and 26.1 eV, which are about 4, 2, and 2% percent larger than those of Li/Na$MnO_2$, Li/Na$CoO_2$, and Li/Na$NiO_2$, respectively. This behavior also follows the localization tendency of the Wannier spread in Fig. 2(b). The Wannier spread is smaller for the transition metals in $XO_2$ than Li/Na$XO_2$ cases. Namely, the localization in $XO_2$ is larger than that in Li/Na$XO_2$, which presumably results from the decreased hybridization channel caused by the absence of Li/Na in the interlayer space.

The screened $U$ values for $XO_2$ systems, however, are smaller than those of Li/Na$XO_2$ systems as in Fig. 2(c). While the increasing tendency from Mn to Ni-system remains, the behavior of the Coulomb parameter among the same TM systems is reversed. The $U$ values for TM-$d$ ions of $MnO_2$, $CoO_2$, and $NiO_2$ are 21, 16, and 12 % smaller than those of Li/Na$MnO_2$, Li/Na$CoO_2$, and Li/Na$NiO_2$, respectively. Another important physics at play here is the screening effect. The ratio of $U/U^{bare}$ effectively quantifies the strength of the screening. Thus, highly reduced $U$ from $U^{bare}$ in Fig. 2(d) indicates the stronger polarizability screening for $XO_2$ systems.

To further analyze the screening effect by O-$p$ orbitals, we have plotted the partial density of states (DOS) of TM-$d$ orbitals and O-$p$ orbitals for all studied systems in Figure 3. We can clearly see that the hybridization of TM-$d$ and O-$p$ orbitals is enhanced upon electron occupation. For example, in the case of Li-based systems, we observe sizable energy separation Mn-$d$ and O-$p$ bands in $LiMnO_2$, which becomes smaller for $LiCoO_2$ case and is eventually closed, and $d$-$p$ bands are well-hybridize for $LiNiO_2$ case, which is described by the systematic increase of TM-$d$ weight in the oxygen originated band energy window from -8 eV to -2 eV. This suggests the larger polarizability screening of later TM cases from the $d$-$p$ channel. Hence, while the differences in $U^{bare}$ are in eV scales (21, 24, and 26 eV for TM in $LiMnO_2$, $LiCoO_2$, and $LiNiO_2$), those for screened $U$ are highly reduced and is of 0.5eV scale (4.28, 4.40, and 4.84 eV for $LiMnO_2$, $LiCoO_2$, and $LiNiO_2$), respectively.

Among the same TM series, the screening is stronger for $XO_2$ systems. Figure 3 shows that for $XO_2$ cases, O-$p$ orbitals move toward the Fermi level compared to Li/Na$XO_2$ systems. Also, TM-$d$ and O-$p$ orbitals are closer in energy indicating the enhanced hybridizations despite the stronger localization in $XO_2$ (See Fig. 2(b)). Eventually, in $XO_2$ systems, the hybridization effects overcome the localization, and the screened $U$ is smaller than that in Li/Na$XO_2$ systems. This demonstrates that complicated physics beyond simple localization picture is involved in the electron-electron interaction.

Furthermore, we have compared the Coulomb parameters for four pairs of polymorphs; O'3/O3-$LiNiO_2$, O3/P2-$NaCoO_2$, O'3/P'2-$NaMnO_2$, and O1/O'3-$NiO_2$. The differences of effective $U$, $U_{eff}$ =$U$-$J$, between the polymorphs are 0.26 eV ($LiNiO_2$), 0.02 eV ($NaCoO_2$), 0.38 eV ($NaMnO_2$), and 0.32 eV ($NiO_2$), which are all less than 0.5 eV as shown in Table S1. Namely, the polymorphs with different layer stacking have similar $U$ values, which indicates $U$ values are mainly determined by a local structure rather than a global structure. The small difference is also originated from the details of the local environment; bond lengths of TM and O in an octahedron. The smallest difference of the average bond length is 0.009 Å in $NaCoO_2$ polymorphs followed by 0.045 Å in $LiNiO_2$, 0.051 Å in $NiO_2$, and 0.104 Å in $NaMnO_2$ as in Table S2. The $U_{eff}$ difference between polymorphs increases as the difference in average bond lengths become larger, which is directly coupled to the hybridization of $d$-$p$ orbitals.

Let us compare our results to the previous studies on Coulomb parameters. Hautier *et al*. reported $U_{eff}$ values to match experimental formation energies, which are widely used in the battery community.[16] Their values of 3.9, 3.4, and 6.0 eV for Mn, Co, and Ni are closer to our calculated values for $Mn^{3+}$, $Co^{3+}$, and $Ni^{3+}$ in Li$XO_2$ than those for $Mn^{4+}$, $Co^{4+}$, and $Ni^{4+}$ in $XO_2$ (Table S1). Zhou *et al*. and Shishkin *et al*. evaluated $U_{eff}$ values self-consistently using the linear response method.[10,21] The $U_{eff}$ values for $LiCoO_2$ and $LiNiO_2$ from Zhou *et al*. (Shishkin *et al*.) are 4.91 (4.87) eV and 6.70 (5.29) eV for $Co^{3+}$ and $Ni^{3+}$, respectively. Overall, the values from the linear response approach are 1.2-2.5eV larger than our values from the cRPA calculations, which affects the calculation of the intercalation voltage as will be discussed later.

Our comprehensive study on cathode systems is a complementary to previous attempts in the *ab initio* calculation of the Coulomb parameters and offers different perspective. Microscopically, we demonstrated that the Coulomb interaction parameters are determined from the various physical origins. To better describe the Coulomb parameters, one should account for the intertwined physics from the electron occupation, screening, and structural details, which are highly system-dependent. Furthermore, our detailed study not only offers reference for Coulomb parameters for the cathode systems, but also paved the way for advanced calculational approaches such as DMFT.

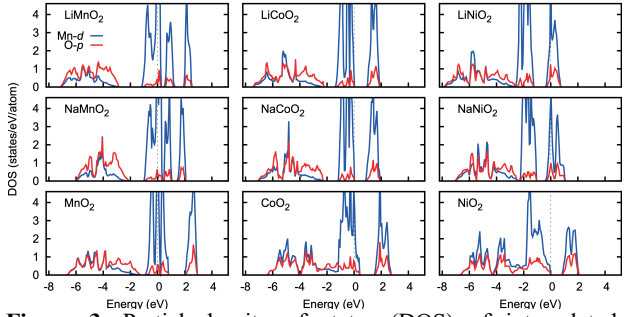

**Figure 3.** Partial density of states (DOS) of intercalated Li/Na$XO_2$ and deintercalated $XO_2$ (X: Mn, Co, and Ni)

The detailed Coulomb interaction parameters $U$ and $J$ are summarized in Table S1. Note that previous approaches for Coulomb parameters reported $U_{eff}$ rather than an explicit evaluation of $U$ and $J$ parameters.[10,16,21] Considering the recent findings of a crucial role of the Hund $J$ in TMOs[53–56] and noting that most of first principles calculations are performed with magnetism on cathode materials, our explicit calculation of both the Coulomb $U$ and Hund $J$ could lead to further studies and better understanding on battery materials.

To demonstrate the validity of our approaches, we applied obtained $U$ parameters to the calculation of the average intercalation voltages for the aforementioned material sets. Figure 4 shows the calculated average intercalation voltage using different functionals; PBE and recently developed meta-GGA functional, SCAN. Three $U_{eff}$ values were considered; $U_{TM^{3+}}$, $U_{TM^{4+}}$ and $(U_{TM^{3+}}+U_{TM^{4+}})/2$. We explicitly included the vdW corrections to consider the weak interlayer interactions.

The calculated voltage with PBE+$U$+vdW using an average $U_{eff}$ value overall agrees well with the experimental values. Since one can adjust the voltage by changing $U$ values without vdW interactions, it might be difficult to find whether the vdW interactions are important or not.[11] With the *ab initio* $U$ value, however, it is clear that the vdW correction is also critical to reproduce the voltage of battery cathode materials. In addition, the results from the PBE calculations underestimate voltages of both Li and Na compounds. We also tested the PBEsol functional for LiCoO$_2$ case, and obtained voltage is similar to that from the PBE functional.

Note that the $U_{eff}$ values from cRPA calculations are 1.2-2.5 eV smaller than the reported $U_{eff}$ values from the linear response method[10,21], and accordingly, the intercalation voltage of 3.79 V for LiCoO$_2$ and 3.52 V for LiNiO$_2$ with PBE+$U$ (averaged $U_{eff}$) calculations are smaller than their or experimental values. It is worth noting that previously calculated intercalation voltages of 3.85 V[13,14] for LiCoO$_2$ and 3.66 V[21] for LiNiO$_2$ using $U_{eff}$ values from the linear response approach are still smaller than experimental voltages. The previous reports and our study indicate that the DFT+$U$ calculations more or less underestimate the average voltage. However, by including the vdW interaction, the intercalation potentials for LiCoO$_2$ and LiNiO$_2$ increase to 4.22 V and 3.94 V, respectively, which agree well with the experimental values of 4.1 V and 3.85 V.[13,57,58] Therefore, our results further show that the inclusion of the vdW interaction as well as an appropriate usage of $U$ is essential for reasonable prediction of the physical properties of layered cathode materials.

SCAN functional (without $U$) also well reproduces the experimental voltage (blue colored dataset in Fig. 4). Recently developed functional, SCAN[32], has been successfully tested for typical TMO systems[59] including the Li-based cathode materials.[11,60] Despite the problems in the description of magnetic systems and the discrepancy in the band gap size with the experimental one[61–63], SCAN known to be a promising methodology which can describe the key physics of transition metal complexes. Hence, we have tested SCAN approaches for Li/Na systems and obtained consistent results with previous reports for Li-ion cathode materials.[11,60] These suggest that SCAN can be an alternative choice for both Li and Na cases when exact enumeration of Coulomb parameters is not available.

The inclusion or local Coulomb correlation over SCAN, SCAN+$U$, in general, overestimates the voltage. As there are still ongoing discussions on the SCAN functional, especially on its capability when cooperated with vdW and +$U$ scheme[11,64–66], the concrete consensus is still to come. Furthermore, note that the hybrid functional overestimates the intercalation voltages of LiCoO$_2$ and LiNiO$_2$ and one needs manually adjust mixing parameter, $\alpha$ to match the experimental values[13,14]. Thus, we claim that PBE+$U$+vdW and SCAN (without $U$) can be the best or the safest choice.

The effect of $U$ and the difference between results from using PBE and SCAN functionals vary upon systems. The voltage of the Co system does not vary a lot no matter how increase $U$ value over 3 eV while the voltages of Ni and Mn systems progressively increase well over 3 eV. The difference in voltages from PBE and SCAN functionals is largest in the Co system (~1 V) followed by the Ni system (~0.7 V), and then the Mn system (~0.4 V). In Mn systems, the results using the SCAN functional is close to the results using the PBE functional with the vdW correction.

By comparing the results of the Li and Na compounds, one can find that the Na compounds show the similar trend with the Li compounds. The calculated and experimental voltages for Na$XO_2$ are lower than those of Li$XO_2$. Interestingly, in NaCoO$_2$ case, the calculated voltages with PBE+$U$+vdW are much larger than the experimental one. The inconsistency between calculation and experiment for NaCoO$_2$ has been reported[15]. Also, note that the estimated voltage in experiment has an error bar toward larger value than 2.8 V denoted by Exp.1 in Fig. 4(d)[15], which makes direct comparison difficult. However, it would be still interesting to check whether other advanced DFT methods, for example, DFT+DMFT can predict the lower voltage for NaCoO$_2$, which can be a further study.

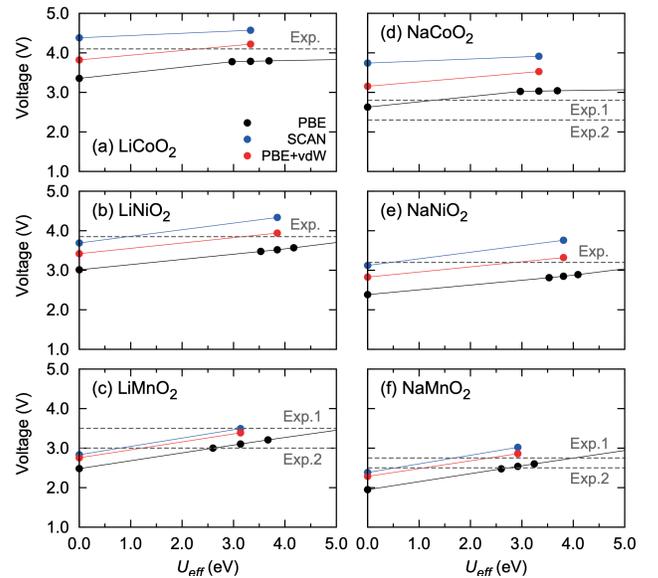

**Figure 4.** The average voltage calculated using several combinations of functionals and vdW correction for different $U$ values. (a) LiCoO$_2$ (b) LiNiO$_2$ (c) LiMnO$_2$ (d) NaCoO$_2$ (e) NaNiO$_2$ (f) NaMnO$_2$. The gray dotted lines indicate the experimental values for LiCoO$_2$[13,57], LiNiO$_2$[13,58], LiMnO$_2$[50,67], NaCoO$_2$[15,68], NaNiO$_2$[69], and NaMnO$_2$[70,71].

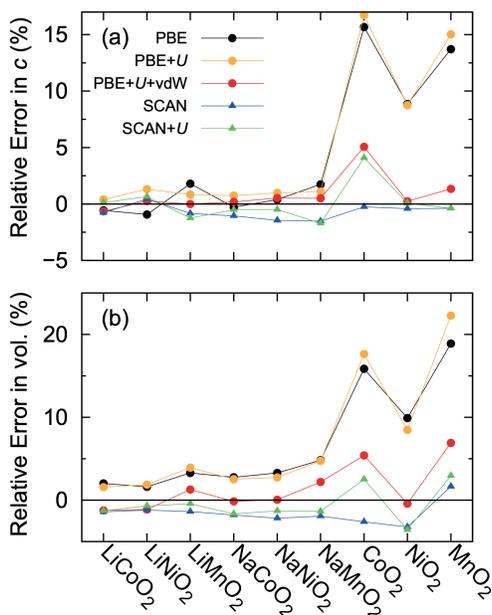

**Figure 5.** Relative errors of (a) the relaxed lattice parameter $c$ and (b) volume/f.u. compared to experiment using PBE, PBE+$U$, PBE+$U$+vdW, SCAN, and SCAN+$U$.

Finally, we compared the relaxed lattice parameters using PBE, PBE+$U$, PBE+$U$+vdW, SCAN, and SCAN+$U$ to the experimental ones. The $U$ values are chosen from our cRPA calculations for each material. Figure 5 shows the relative errors of lattice parameter $c$ and volume/f.u. of fully relaxed structures. The exact numbers for calculated and experimental lattice parameters are tabulated in Table S3. Overall, the results of PBE+$U$+vdW calculations agree well with experimental ones of intercalated and de-intercalated structures while the PBE(+$U$) calculation generally overestimates lattice parameters while the SCAN(+$U$) underestimates.

PBE and PBE+$U$ calculations have similar results meaning that just adding $U$ does not much improve much the reproducibility for the lattice parameters in PBE functional. An inclusion of the vdW interaction, however, is critical to reproduce the interlayer distance (denoted by lattice parameter $c$) and volume with the PBE functional, especially for the de-intercalated materials $X$O$_2$, whose interlayer interaction is the vdW type. The interlayer distance is also well reproduced using SCAN functional even without vdW correction. Adding $U$ alone does not affect the results much for lattice parameters, not like the voltage case, but we can still conclude that PBE+$U$+vdW well describe the physical properties; average intercalation voltage and lattice parameters of layered materials at the same time.

## 4. CONCLUSIONS

In conclusion, we evaluate the Coulomb interaction parameters $U$ and $J$ self-consistently and demonstrate that choosing *proper U* is essential for the explanation of physical properties; average intercalation voltage and lattice parameters in layered Li- and Na-ion cathode materials. The Coulomb interaction parameters in cathode materials are affected by intertwined physics such as electron localization and screening. The calculated $U$ values for Li compounds are close to those of Na counterparts. The polymorphs with different layer stacking also have similar $U$ values showing that the important factor for $U$ determination is a local environment rather than a global structure. By self-consistently calculating $U$ values, our PBE+$U$+vdW calculations well reproduce the experimental intercalation potential and lattice parameters without any external parameters. Also, it shows that both appropriate $U$ and the vdW interaction are critical to investigate the physical properties of Li- and Na-ion cathode materials. Despite the SCAN functional without $U$, which also does not need an external parameter, provides the comparable results with PBE+$U$+vdW, knowledge on the $U$ values for each system can offer the deeper understanding on the system such as strength of the electron localization and hybridizations. Using that, one can establish a simplified but intuitive model to understand the underlying mechanism.

Furthermore, it is worth noting that the original modeling of the initial Li-ion battery system by Goodenough is based on the simple model with representative physical parameters[1–3] from experimental results and intuitions. This shows that the correct identification of the related parameters such as charge transfer energy, crystal field energy, bandwidth, and Coulomb parameters obtained within *ab initio* description can act as a fundamental step for designing the new materials. Thus, we believe that this systematic study can act as a bridge between the DFT and model-based approaches and, by *ab initio* calculation of Coulomb parameters, not only we can expect predictive power for first-principles calculations but also offer the inputs for model-based simulations for designing advanced battery materials.

## ASSOCIATED CONTENT

### Supporting Information

The Supporting Information is available free of charge on the ACS Publications website.

The Wannier-projected bands with DFT bands; Self-consistently evaluated $U$, $J$ by the cRPA approach; Experimental TM-O distances in an octahedral; Calculated lattice parameters with different computational condition (PDF)

## AUTHOR INFORMATION

### Corresponding Author

Sooran Kim - *Department of Physics Education, Kyungpook National University, Daegu 41566, Korea*; https://orcid.org/0000-0001-9568-1838; Email: sooran@knu.ac.kr

### Authors

Bongjae Kim - *Department of Physics, Kunsan National University, Gunsan 54150, Korea*; https://orcid.org/0000-0002-3841-2989

Kyoo Kim - *Korea Atomic Energy Research Institute (KAERI), 111 Daedeok-daero, Daejeon 34057, Korea*; https://orcid.org/0000-0002-7305-8786

### Notes

The authors declare no competing financial interest.

## ACKNOWLEDGMENT

We thank the helpful discussion with Dong-Hwa Seo and Dan T. Major. This work was supported by NRF [Grant No. 2019R1F1A1052026, No. 2018R1D1A1A02086051, 2016R1D1A1B02008461], KISTI Supercomputing Center [Project No. KSC-2019-CRE-0172, KSC-2019-CRE-0231] and Internal R&D program at KAERI (Grant No. 524210-20).

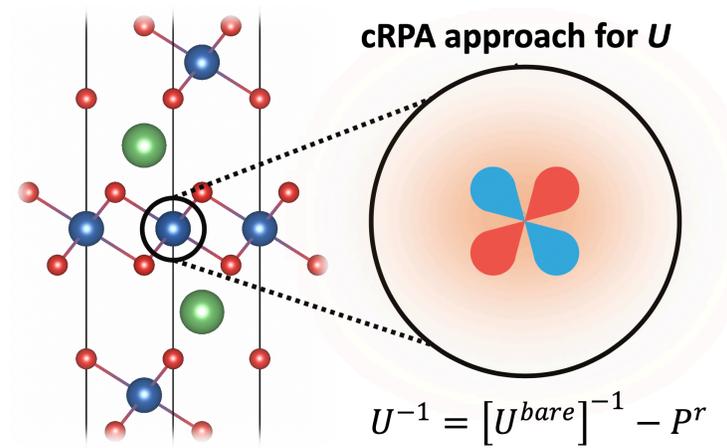

Table of Contents artwork